# Beyond DNS: Unlocking the Internet of AI Agents via the NANDA Index and Verified AgentFacts

Ramesh Raskar (MIT), Pradyumna Chari (MIT), John Zinky (Akamai), Sichao Wang (CISCO), Rekha Singhal (TCS), Robert Lincourt (Dell), Mahesh Lambe (UnifyDynamics), Jared James Grogan, Rajesh Ranjan (CMU), Shailja Gupta (CMU), Raghu Bala (Synergetics AI), Aditi Joshi, Abhishek Singh (MIT), Ayush Chopra (MIT), Dimitris Stripelis (Flower AI), Bhuwan B (HCLTech), Sumit Kumar (HCLTech), Maria Gorskikh

*Project NANDA*

**Abstract:** The Internet is poised to host billions to trillions of autonomous AI agents that negotiate, delegate, and migrate in milliseconds and workloads that will strain DNS-centred identity and discovery. In this paper, we describe the NANDA index architecture, which we envision as a means for discoverability, identifiability and authentication in the internet of AI agents. We present an architecture where a minimal lean index resolves to dynamic, cryptographically verifiable AgentFacts that supports **multi-endpoint routing, load balancing, privacy-preserving access, and credentialed capability assertions**. Our architecture design delivers five concrete guarantees: (1) A quilt-like index proposal that supports both NANDA-native agents as well as third party agents being discoverable via the index, (2) **rapid global resolution** for newly spawned AI agents, (3) **sub-second revocation and key rotation**, (4) schema-validated capability assertions, and (5) privacy-preserving discovery across organisational boundaries via verifiable, least-disclosure queries. We formalize the AgentFacts schema, specify a CRDT-based update protocol, and prototype adaptive resolvers. The result is a lightweight, horizontally scalable foundation that unlocks secure, trust-aware collaboration for the next generation of the Internet of AI agents, without abandoning existing web infrastructure.

## I. INTRODUCTION

Autonomous AI agents are emerging as a critical abstraction layer responsible for executing tasks, communicating, reasoning, and making decisions on behalf of human and machine stakeholders. From price-negotiation bots optimizing supply-chain contracts, customer-service agents that rebook millions of flights and digital finance, healthcare orchestration and scientific collaboration, AI agents are expected to scale into trillions of active participants in a globally distributed, interconnected environment.

This evolving reality demands foundational infrastructure that supports privacy, discovery, low-latency resolution, and verifiable trust evaluation of AI agents across trillion-scale, decentralized and highly dynamic ecosystems. While DNS and HTTPS certificates excel at mapping human-readable domains to static endpoints, their minute-to-hour update cycles and ownership-only trust model cannot satisfy the millisecond orchestration, rapid revocation, and code-attestation needs of agent-centric systems. In this paper, we propose an architecture for a lean NANDA index, which when paired with an AgentFacts schema (self-describing bundles of AI agent identity and capabilities) enables resolution that is dynamic (< 1 s global propagation), privacy-preserving, low-latency and without compromising extensibility, interoperability, and trust. Our primary contributions are as follows:

1) We define a **minimal core index** that holds only essential static metadata (agent IDs, credential pointers, and dynamic agent facts URLs) e.g. ≤ 120 bytes per record, to reduce update needs. We also introduce a **self-describing and verifiable AgentFacts schema i.e.** JSON-LD documents whose @context enables forward-compatible schema evolution that allows agents to dynamically update capabilities, endpoints, and authentication info without modifying the index. **TTL-based endpoint resolution** supports decoupling, enabling caching, DDOS protection, and flexible deployment.
2) Critically, the index is designed from the bottom-up for **interoperability with enterprise agent registries**, allowing for a range of ways in which agents can be made visible via the NANDA index.
3) **Dual-Path Privacy Resolution mechanism** enables metadata lookups without revealing the requester's identity, aligning with zero-trust principles. Additionally, **AgentFacts enables cryptographic verification** from trusted whitelisted issuers to ensure authenticity and prevent tampering; short-lived verifiable credentials (< 5 min) allow sub-second revocation.

At a high level, the index transforms the complexity of a fully connected N×N network, where every agent must maintain direct links with all others, into a simpler 2N problem by acting as a shared handshake channel. This allows any two agents to discover and connect through a common interface. Once the initial handshake is complete, the index is no longer required for ongoing communication between those agents, unless the session expires based on a defined time-to-live (TTL) parameter.

The current Domain Name System (DNS), foundational to the internet since 1983, was designed for static web infrastructure, not for the dynamic, context-aware operation of autonomous AI agents. As we envision a future with billions or even trillions of interoperable agents, DNS becomes a bottleneck, not just in terms of scale but in its architectural assumptions. We argue that the emergence of an open, agentic internet cannot be realized within the constraints of DNS. To enable this shift, we propose a new system that decouples identity from metadata, supports multi-hop resolution, and provides a flexible substrate for agent discovery, routing, and capability negotiation. This infrastructure is not merely a scaling solution, it is a prerequisite for the agentic web itself.

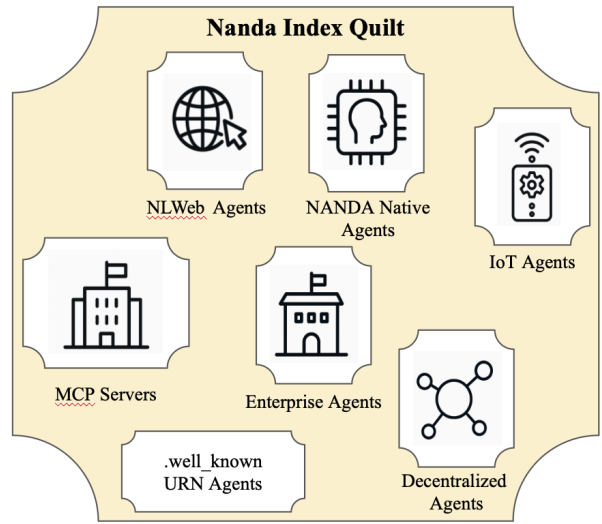

**Figure 1: The NANDA index is a quilt of various agent, resource and tool registries.**

We view the NANDA index as a quilt of agents, resources, and tools across platforms, organizations and protocols. Through such an approach, we allow for global interoperability, discoverability, and flexible governance of agents: NANDA need not authenticate, authorize and govern all agents that exist on the internet of AI agents. Commercial, governmental and individual entities may determine whether agents are directly certified and visible via the NANDA index, or if the NANDA index merely holds a redirect to their platforms that allows them to have deeper control to agents. Figure 1 highlights this vision.

## II. DESIGN GOALS

The design of the Internet-scale decentralized AI agent index infrastructure must account for a wide spectrum of operational and architectural requirements. These range from performance and scalability to security, privacy, trust, governance and resilience. In this section, we outline the key goals that informed the proposed index model.

**Problems we are solving [7]:**

- **Mitigate Index Writes & Delay**: DNS, BGP, and WHOIS were built for millions of static records, not **billions of updates per hour** or sub-second worldwide sync.
- **Trust Gap**: A TLS cert only proves you own a domain; it says nothing about an agent's code or behaviour. We need **signed capability proofs and instant revocation**.
- **Privacy & Split-Horizon** (i.e. returning different answers to the same name-lookup depending on who is asking or where the query originates): Today's look-ups expose who is asking for what. Enterprises need **private, split-horizon resolution and tamper-evident audit logs**.
- **Routing Limits**: Fixed A/AAAA records can't keep up with agents that move every few seconds or require geo-based DDoS shuffling.
- **Governance & Liability**: Without a transparent, append-only log, no one can prove compliance or assign blame when agents misbehave across borders.

Before we dive into the individual design goals, it is useful to explain **how the raw pain-points enumerated above translate into the specific targets that follow**. Our approach is deliberately reductionist:

- **Isolate the bottleneck.** For each problem [7], ask *what single property of the current stack makes it untenable at the trillion-agent internet scale?*
- **Introduce exactly one constraint-shifting primitive**, e.g., lean-record, privacy path, or VC-signed fact that removes that bottleneck without disturbing layers that already scale.

- **Accept and surface the trade-off.** Every fix (latency, privacy, richness) carries an equal-and-opposite cost elsewhere; listing pros/cons next to each goal keeps those tensions explicit.

The goals that follow form an **orthogonal checklist** rather than a monolithic blueprint: an implementer may adopt Goal A and Goal F first, defer Goal D if latency is paramount, or swap Goal E's adaptive resolver for an in-house service mesh. The score-card on the next page summarises how each goal closes a problem, what it gives us, what it costs, and which goals it may collide with; the subsections that follow provide the technical detail.

### A. Lightweight Reference Index

**Goal:** The index must achieve lean, stable operation with records constrained to ≤120 Bytes to ensure scalability and fault tolerance across distributed agent networks.

**Issues:** Current DNS infrastructure processing 10 Billion daily lookups for only 300M addresses demonstrates that direct scaling approaches with dynamic agent state would be unsustainable for billions of unique agents, creating excessive write overhead and infrastructure complexity.

**Proposal:** Implement the index as a static reference layer that maps agent identifiers to endpoints plus relevant metadata, reducing write frequency by ~$10^4\times$ relative to DNS while lowering infrastructure complexity and improving fault tolerance.

**Co-dependencies:** This architecture requires strict separation between static routing information handled by the index and dynamic agent state managed independently at the endpoints.

**Future Evolution:** The minimal core design may need to accommodate additional metadata fields or enhanced routing mechanisms while preserving the fundamental static reference mapping principle to maintain performance advantages over dynamic index systems.

### B. Enabling Diverse Agent Registration Models

**Goal:** An index architecture that accommodates diverse agent registration models while maintaining privacy, split horizon governance, and liability management across heterogeneous administrative domains.

**Issues:** Monolithic index designs cannot address the varied governance, privacy, and liability requirements of agents operating under different frameworks.

**Proposal:** Deploy a quilt-like architecture enabling multiple registration types: agents can be natively registered directly via the NANDA index or administered by external commercial/governmental entities while maintaining visibility through the NANDA index interface, as detailed in Table 1.

**Co-dependencies:** This model requires interoperability protocols between NANDA and external registries, along with consistent metadata standards across boundaries.

**Future Evolution:** The quilt architecture may expand to support additional registration categories and cross-index federation mechanisms while preserving the core principles of distributed governance and liability separation.

| Entity | Agent Name | Link to Agent and AgentFacts | Comment |
|---|---|---|---|
| Civil Society | @agentx | NANDA native URL | Agents for individuals, or unaffiliated commercial entities |
| Government | @japan:shop | NANDA native URL | Location-specific domains, similar to .com, .co.uk etc. |
| Enterprise (Agent-stores) | @company | Routed via @company | Agents within enterprise agent store (Salesforce, Google, Microsoft, Amazon etc.) but only accessible via first going to the agent store |
| Enterprise (Agent-stores) | @company:shop | URL administered by 'company' | Agents within enterprise agent store (Salesforce, Google etc.) but also directly visible on the index |
| Web3 | @DID:company | Routed via company marketplace | Agents within Web3 agent markets, only accessible via first going to the agent store |
| Web3 | @DID:company:agent | DID administered by company, routed via NANDA | Agents administered by Web3 agent markets, and authenticated via decentralized identifiers |

**Table 1:** The NANDA index is envisioned as a quilt of directly-registered agents, but also enterprise and governmental agents visible on the index to varying degrees.

**C. Endpoint Agility and Sub-Second Reachability**
**Goal:** Dynamic endpoint management for AI agents operating in variable deployment environments requiring frequent network changes during blue/green deployments, geo-failover, and traffic-shaping operations while maintaining sub-second reachability.
**Issues:** Static endpoints create scalability bottlenecks and security vulnerabilities including DDoS attack susceptibility, while frequent index updates for endpoints would violate the minimal core architecture principles.
**Proposal:** Decouple endpoint resolution from the index by delegating it to AgentFacts documents containing TTL-scoped endpoint lists: static (1-6h), rotating (5-15 min), or adaptive (30-60s), with the index/enterprise registries storing FactsURL/PrivateFactsURL references and AgentFacts signatures covering endpoint lists to prevent spoofing.
**Co-dependencies:** This architecture requires TTL parameter support for adaptive caching, cryptographic signature validation for AgentFacts documents, and coordination between index lookups and facts resolution.
**Future Evolution:** The endpoint agility system may incorporate more granular TTL categories, enhanced load balancing algorithms, and advanced security mechanisms..

**D. Improving Scalability through Decentralizations**
**Goal:** Asynchronous metadata publishing and revision by agents or publishers without index write-overheads, supporting 10k updates/sec per index shard.
**Issues:** Centralized metadata updates create bottlenecks and dependencies on index availability, limiting system scalability and agent autonomy in dynamic environments.
**Proposal:** Implement decentralized updates where agents sign AgentFacts JSON-LD documents with W3C Verifiable Credentials, upload to self-selected hosts (IPFS CID, CDN, or enterprise bucket), and publish the resulting URL under existing facts_url pointers in the lean index, enabling hosting of AgentFacts without index writes.
**Co-dependencies:** Requires Verifiable Credential infrastructure, cryptographic signature validation, and coordination between multiple hosting platforms for facts.
**Future Evolution:** May expand to support additional credential standards and hosting mechanisms while maintaining the core principle of decentralized, authenticated metadata management.

**E. Privacy Preservation**
**Goal:** Privacy-preserving agent resolution aligned with data minimization principles including purpose limitation and unlinkability, preventing exposure of requester identity (IP, AgentID, capability queries) except timing leaks.
**Issues:** Standard resolution exposes access patterns to passive tracking by agents, network observers, and malicious index/registry operators, violating privacy best practices for sensitive workloads.
**Proposal:** Deploy dual-path resolution via PrimaryFactsURLand PrivateFactsURL , where PrivateFactsURL can be hosted by third parties or decentralized storage to shield access patterns, with cacheable requests or mix-net broadcasts preventing traffic-analysis correlation and AgentAddr signatures covering both URLs to prevent endpoint swapping attacks.
**Co-dependencies:** Requires third-party hosting infrastructure, protocols for anonymous requests, cryptographic signature validation, and policy frameworks.
**Future Evolution:** May incorporate advanced anti-timing attack mechanisms and enterprise-specific "PrivateFacts-only" modes while balancing privacy protection with latency requirements.

**F. Flexible Routing Choices**
**Goal:** Provide every AI agent with latency-, privacy-, and resilience-aware routing choices analogous to Akamai's Global Traffic Management service for real-time optimal endpoint selection.
**Issues:** Single-path resolution cannot accommodate diverse operational requirements including geo-distributed load balancing, failover routing, DDoS shuffle-sharding, and capability-specific dispatching across varying network conditions.
**Proposal:** Implement three routing mechanisms: static endpoint discovery via PrimaryFactsURL , privacy-preserving metadata fetch via IPFS/Tor-style relays as PrivateFactsURL , and cryptographically signed AdaptiveresolverURL for programmable real-time endpoint selection included in AgentAddr or AgentFacts.
**Co-dependencies:** Requires IPFS/Tor relay infrastructure, real-time routing algorithms, cryptographic signature validation for adaptive resolvers, and coordination between static and dynamic endpoint selection mechanisms.
**Future Evolution:** May incorporate machine learning-based routing optimization and enhanced shuffle-sharding algorithms while maintaining the core principle of flexible, multi-criteria path selection for diverse operational scenarios.

**G. From Self-Advertising to Audited Metadata**
**Goal:** Establish verifiable claims infrastructure preventing AI agents from spoofing capabilities, impersonating reputable actors, or conducting supply-chain poisoning, sybil attacks, and traffic diversion.
**Issues:** Self-advertising metadata enables malicious agents to make unverifiable claims about capabilities and identity, creating security vulnerabilities and trust deficits.
**Proposal:** Treat AgentFacts as verifiable claims requiring W3C Verifiable Credential v2 cryptographic attestation, with issuers including enterprises, consortiums, or federated certification authorities operating within domain-specific trust zones that can cross-sign each other, and each claim linked to credentialing paths anchored in issuer IDs with revocation status via VC-Status-List.

**Co-dependencies:** Requires W3C Verifiable Credential infrastructure, ID resolution systems, cross-signing protocols between trust zones, and VC-Status-List revocation mechanisms.

**Future Evolution:** May expand trust zone federation models and enhanced revocation mechanisms while maintaining the core principle of cryptographically attested, auditable agent metadata for decentralized, extensible, and trustworthy discovery infrastructure.

Table 2 summarises how each design goal (A,G) closes a pain-point highlighted in the companion "Why" paper, then rates the solution's current maturity. Green ticks (✔) denote fully specified features in the present draft; yellow triangles (△) indicate partial support; red crosses (✖) remain open research items..

| Goal | Core problem solved | Key pros | Key cons | Status* |
|---|---|---|---|---|
| A: Lightweight Reference Index | DNS write-overhead & propagation lag | $10^4 \times$ fewer writes | Lean schema limits rich search | ✔ |
| B: Diverse Agent Registration Models | Allowing varying degrees of enterprise autonomy over agents | Quilt of existing agent registries that can be accommodated within NANDA | Lower quality of information about highly controlled agents, within the NANDA index/AgentFacts | ✔ |
| C: Endpoint Agility and Sub-Second Reachability | Endpoint churn every few seconds | < 1 s fail-over / DDoS shuffle | Resolver cache churn | △ |
| D: Improving Scalability through Decentralization | Index bottleneck at 10 B updates / h | Unlimited parallel self-hosted updates | Replay window risk | ✔ |
| E: Privacy Preservation | Query privacy & split-horizon governance | Unlinkable look-ups / least-disclosure | Adds 30-60 ms latency | △ |
| F: Flexible Routing Choices | Geo-LB, fail-over, capability dispatch | Operator chooses best path | Client-side complexity | ✔ |
| G: From Self-Advertising to Audited Metadata | Capability spoofing & supply-chain attacks | Offline VC verification | VC payload bloat | ✔ |

* **Status legend** – ✔ fully specified, △ partial / open items, ✖ design stub.

**Table 2: Summary of each design goal, the problem it solves, pros, cons, and current maturity**

Table 2 maps each design goal to its motivating problem statement, highlights residual risks, and surfaces conflicts we resolve in Sections IV,VI.

**III. SYSTEM ARCHITECTURE OVERVIEW**

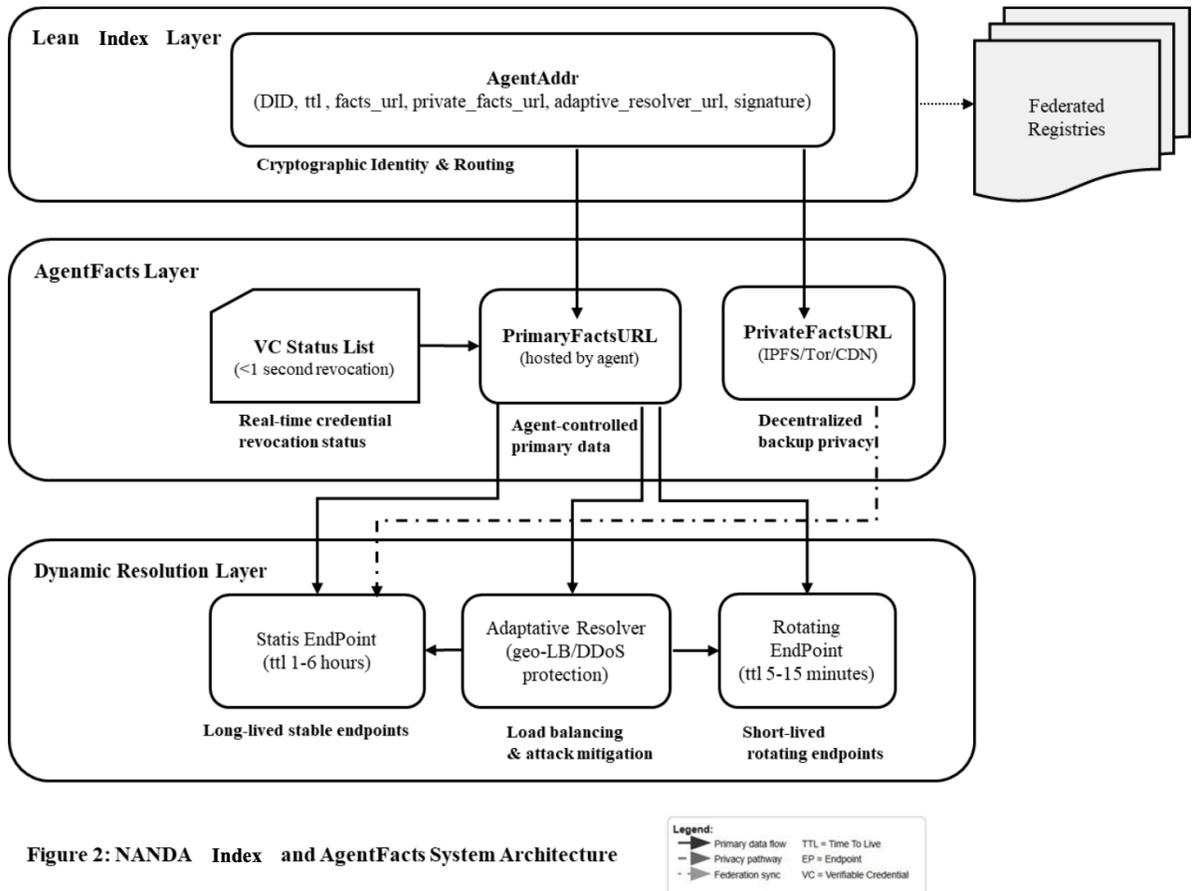

Figure 2: NANDA Index and AgentFacts System Architecture

The architecture of the proposed index system is organized into a modular, hierarchical stack designed to support internet-scale, autonomous AI agent ecosystems. At its core, the system cleanly separates (1) static identity resolution, (2) verifiable metadata distribution and (3) dynamic endpoint routing. This separation enables both performance optimization and secure, decentralized operation across heterogeneous networks e.g., cuts index write load by $10^4\times$ while keeping <1 s global resolution.

To meet the stringent requirements of **low-latency discovery, cryptographic trust evaluation, privacy, and cross-domain governance**, we structure the architecture into a hierarchy with **three levels: Lean Index, AgentFacts, and Dynamic Resolution**, each optimised for a distinct function. Figure 2 highlights this architecture.

1. **Index Level (Anchor Tier) :**
   ○ Provides decentralized mapping from agent identifiers (IDs) to metadata URLs i.e. **AgentAddr** ($\leq$ 120 B) containing TTL and pointers (FactsURL, PrivateFactsURL, AdaptiveResolverURL).
   ○ Records are cacheable and Ed25519-signed, slashing index-write overheads while blocking tampering. Maintains only essential, static metadata such as Agent ID, credential hashes, TTL, and URLs for dynamic agent facts.
   ○ Supports resolution via short-lived, cacheable records that are resistant to tampering and unauthorized updates.

2. **AgentFacts Level (Metadata Distribution Tier):**
   ○ Hosts self-describing JSON-LD documents (AgentFacts), each signed as a W3C VC., that contain endpoint lists, capability descriptors, telemetry configurations, authentication protocols, and credentialed evaluations.
   ○ Enables frequent, independent updates without requiring index-level intervention.

  - Without touching the index, it supports hosting at either agent-owned domains (via .well-known or IPFS) or neutral third parties to support privacy-preserving resolution.

3. **Dynamic Resolution Level (Adaptive Routing Tier):**
   - Dynamically interprets AgentFacts metadata to resolve live endpoints (TTL 1-6 h), rotating pool (TTL 5-15 min), or apply adaptive routing policies, and maintain privacy constraints.
   - Includes support for decentralized adaptive resolvers that can load-balance, geolocate, or behavior-route traffic as needed.
   - TTL-based resolution ensures endpoint freshness while minimizing re-resolution overhead.

This layered model allows the core index to remain minimal and stable, while richer metadata and operational agility are pushed to the agent facts and resolution services.

**IV. THE LEAN INDEX**

The index resolver does not return raw metadata or service endpoints. Instead, it returns a **cryptographically signed, cacheable object** called AgentAddr, which serves as a lightweight address record for the agent. This object encapsulates the agent's identifier, TTL (time-to-live), and pointers to metadata and optional routing infrastructure.

- **agent_id** *(Machine-readable ID))*: A globally unique agent identifier
- **agent_name** *(URN)*: A human-readable alias encoded as a URN (e.g., urn:agent:salesforce:starbucks).
- **primary_facts_url** *(PrimaryFactsURL)*: A reference to the AgentFacts hosted at the agent's domain (e.g., https://salesforce.com/starbucks/.agent-facts).
- **private_facts_url** *(PrivateFactsURL)*: An optional, privacy-enhanced reference to the AgentFacts hosted on a third-party or decentralized service (e.g., https://agentfactshost.com/...).
- **adaptive_resolver_url** *(optional)*: An optional endpoint for dynamic routing services (e.g., https://resolver.example.com/dispatch) that handle load balancing, failover, or geo-aware dispatching.
- **ttl** *(Time-To-Live)*: The maximum cache duration before the client must re-resolve the record.
- **signature**: A cryptographic signature from the index resolver that binds the contents of the AgentAddr.

The AgentAddr object is **signed and cacheable**, allowing clients to redistribute it and perform verification without repeated lookups. Its role is similar to a DNS record, but extended to include verifiable metadata pointers and flexible routing instructions.

We show an example entry in Table 3 below, of an agent owned and governed by company, but visible via the NANDA index.

| Field | Value |
| --- | --- |
| agent_id | nanda:550e8400-e29b-41d4-a716-4466554400 |
| agent_name | agent:Company:TranslationAssistant |
| primary_facts_url | https://TranslationAssistant.salesforce.com/.agent-facts |
| private_facts_url | https://agentfactshost.com/550e8400-e29b-41d4-a716 |
| adaptive_resolver_url | https://resolver.salesforce.com/dispatch/translation |
| ttl | 3600 (1 hour) |
| signature | signature_hash_placeholder |

**Table 3: An example entry in the NANDA quilt of registries.**

**A. Lean Index Resolution Paths**

The index supports multiple agent resolution workflows depending on context, privacy requirements, and routing logic. These workflows begin with resolution of the AgentName into an AgentAddr, which guides the next step.

1. **Direct Communication (Endpoint Access):** This path is used when the client intends to connect to an agent's endpoint directly and minimal metadata is required. This is used when privacy is not a concern and agent-hosted metadata is available.

AgentName → Index → AgentAddr → Endpoint

**Metadata Resolution (for trust or discovery) can be** used when the client wants to discover the agent's declared capabilities, audit results, or telemetry configuration via its signed AgentFacts. Metadata may be used for interest validation, trust scoring, or reputation checks. Endpoint access is optional and performed only after validation.

AgentName → Index → AgentAddr → FactsURL or PrivateFactsURL → AgentFacts

2. **Enterprise Registry Resolution:** This path is used when the client intends to connect to an agent that is hosted within an enterprise registry. If the NANDA index only holds a reference to an enterprise registry (either a centralized or a DID-based registry), the client is routed to the enterprise registry, from where they can be routed either to the AgentFacts or the agent endpoint per their requirement.

   AgentName → Index → EnterpriseRegistry → FactsURL or PrivateFactsURL or Endpoint

3. **Privacy-Preserving Resolution via PrivateFactsURL:**
   When the client wishes to preserve its identity or avoid contacting the agent's infrastructure, it uses the PrivateFactsURL. This path does not resolve to an endpoint, and endpoint engagement is a separate decision, possibly through proxy or post-authentication.

   AgentName → Index → AgentAddr → PrivateFactsURL → AgentFacts (metadata only)

4. **Adaptive Routing Resolution:** When the agent supports dynamic, context-aware routing (e.g., for load balancing, geo-awareness, or DDoS protection), the index returns the resolver URL directly in the AgentAddr. The AdaptiveResolver may return temporary signed endpoints or session tokens. The AgentFacts is not required unless client policies demand additional validation.

   AgentName → Index → AgentAddr → AdaptiveResolver → (Ephemeral) Endpoint

These resolution paths are unified by the use of cryptographically signed AgentFacts and credential validation logic that ensures agents cannot falsify capabilities or impersonate others. Figure 3 describes the resolution flow steps.

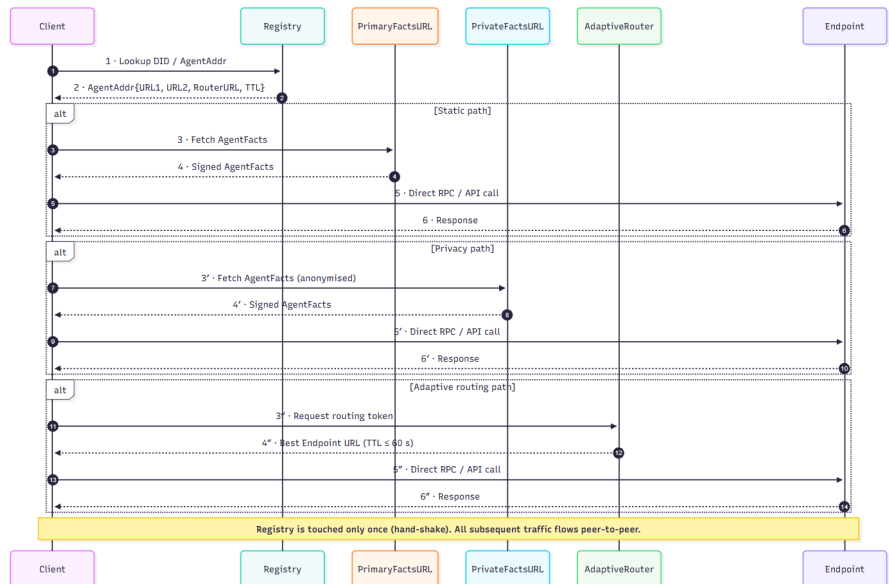

Figure 3: The resolution flow steps on the NANDA index architecture.

**B. Quilt-Like Index Model**
The index can be deployed as a quilt of registries to support scalability across sectors and jurisdictions. Each index entry may:
- Point to an agent registered natively via the NANDA index
- Point directly to an enterprise agent registry, which will handle further agent resolution

- Point to an enterprise agent whose identity and verification is handled by the relevant enterprise registry, but is directly registered as part of the NANDA index.

Each agent may also be identified by a traditional identifier that is centrally certified, or a decentralized identifier. Table 1 describes architecture, with the goal being to maximize interoperability with enterprise registries, verification and trust mechanisms and enabling enterprise players to decide the level of autonomy they desire while being part of the open internet of AI agents. Through this quilt-like approach to hosting the index, the we are able to support a range of agents, resources and tools: MCP server registries [9], Google A2A native agents and registries [8], Microsoft NLWeb agents [10], as well as covering a range of communication protocols for AI agent interoperability [8, 9, 11, 12, 13, 14, 15].

### C. Distributed Index Model

We also envision for the index to be deployed as a globally distributed system. This allows for redundancy and load distribution. Each index instance may:
- Be operated by an enterprise, cloud provider, or neutral public body.
- Federate with other registries using verifiable links, credential exchanges, or trust frameworks.

This federated model ensures organizational autonomy, fault tolerance, scalability and interoperability, crucial for a global AI agent network.

## V. AGENTFACTS SCHEMA AND RESOLUTION MECHANISM

The AgentFacts is a structured, cryptographically signed metadata document that encapsulates an AI agent's dynamic state, declared capabilities, endpoints, and trust-related credentials. It plays a central role in decoupling the lean index from volatile or sensitive metadata, enabling agile agent discovery and interaction at scale. This section formalizes the AgentFacts schema and describes its resolution pathways.

### A. Role and Benefits of the AgentFacts

The AgentFacts functions as an intermediary data structure between a stable index record and an agent's operational configuration. It offers the following advantages:
- **Dynamic Updatability:** AgentFacts can be updated independently of the index, supporting frequent changes to capabilities, telemetry, or endpoints.
- **Credentialed Assertions:** All claims (e.g., skills, certifications) are signed by issuers, ensuring tamper-resistance and reputation integrity.
- **Deployment Flexibility:** Facts can be hosted at the agent's domain (**PrimaryFactsURL**) or on neutral infrastructure (**PrivateFactsURL**) to meet privacy or reliability requirements.
- **Multi-Endpoint Support:** Enables load balancing, routing to specialized instances, and failover mechanisms.

AgentFacts serve as the functional equivalent of a decentralized service manifest. Each facts includes:

- Versioned metadata and labels (e.g., "Translation Assistant v1.2.1").
- Lists of capabilities (e.g., "streaming", "authentication").
- Skills and modalities (e.g., "text → speech" translation).
- Telemetry endpoints for observability (e.g., via OpenTelemetry).
- Signed evaluations and certifications.
- Endpoint URIs and resolution TTLs.

All AgentFacts must be signed by an issuer authorized under the NANDA trust framework or domain-specific federated authority.

**Why another metadata artefact?**
The agentic community already has descriptors such as Google's *A2A Agent Card*, as well as the concept of agent metadata descriptors (Grogan, 2022-2025). Before detailing our own schema, it's useful to show what the A2A card solves and equally, what it leaves unsolved at the trillion-agent scale. Table 4 makes that contrast explicit.

| Dimension | Google A2A Agent Card | AgentFacts (this paper) |
|---|---|---|
| **Purpose** | Advertise a server-side agent's endpoint and skills for JSON-RPC interaction | Convey live endpoints **plus** cryptographically-signed capabilities, SBOM hash, privacy path, revocation info |
| **Discovery Path** | One-step fetch (due to absence of index piece) at /.well-known/agent.json | **Two-step**: lean index → FactsURL / PrivateFactsURL |
| **Endpoint Freshness** | Assumes minutes-to-hours stability (no TTL field) | TTL-scoped lists: static (1-6 h) or rotating (5-15 m) |
| **Trust Primitive** | Plain HTTPS and optional token, self-declared attributes | W3C Verifiable Credential v2 signatures; VC-Status revocation (<1 s), support for third-party audited attributes |
| **Privacy Option** | None (lookup hits agent host) | Optional **PrivateFactsURL** via IPFS/Tor; hides requester |
| **Schema Weight** | ≈ 0.3–1 KB JSON | 1–3 KB JSON-LD + VC |
| **Best-fit Use Case** | Stable SaaS agents inside a single marketplace | Highly mobile, privacy-sensitive, or safety-critical agents at trillion scale |

**Table 4: AgentFacts vs Google A2A Agent Card**

**Take-away:** an A2A card is ideal for a single-marketplace, server-hosted agent whose endpoint changes infrequently. AgentFacts adds four capabilities that our target workloads demand: (1) TTL-scoped endpoint lists for minute-level redeploys, (2) verifiable credentials for code-integrity and capability proofs, (3) a privacy path (PrivateFactsURL) for split-horizon governance, and (4) millisecond revocation via VC-Status lists. The remainder of Section IV specifies how those features are encoded in JSON-LD and resolved in < 1 s end-to-end.

### B. Formal Schema Specification

AgentFacts are represented in JSON-LD and conform to a versioned schema. A minimal example is shown in Table 5 below, and the full schema can be found in the appendix.

| | **Description** |
|---|---|
| **ID:** | Unique identifier |
| **AgentName:** | Human readable name |
| **Endpoints:** | Can be URL or DID, links to various hosted versions of the service, or even a dynamic routing service |
| **UsageFormat:** | Input output formats, API spec, protocol support, auth |
| **Certification:** | Certified via a index or self-certified via DID |
| **Capabilities:** | Declared capabilities, performance on those capabilities, with the option for third-party auditing and certification |
| **Discovery:** | Metadata for search |
| **Security:** | Security requirements, including optional third party security auditing |

**Table 5: A minimal representation of the AgentFacts schema.**

This schema includes both static metadata (e.g., identifiers, descriptions) and dynamic fields (e.g., capability metrics, endpoint lists), all bound by verifiable cryptographic signatures.

Accordingly, AgentFacts can be viewed as a superset of the Agent Card: any conforming A2A server can embed its existing card as a skills extension, gaining cryptographic attestation, privacy paths, and TTL-based routing without altering its runtime logic

### C. AgentFacts Hosting Patterns
AgentFacts can be resolved through two primary URL schemes:

- **PrimaryFactsURL: Primary Facts Hosting**
  Hosted under a standardized path on the agent's domain (e.g., https://example.com/.well-known/agent-facts). It is specified as facts_url in the AgentAddr object, recommended for agents operating public-facing services or enterprise-controlled deployments, and enables clients to retrieve metadata directly from the source domain. "facts_url": "https://salesforce.com/starbucks/.agent-facts"

- **PrivateFactsURL: Privacy-Preserving Hosting (Third-Party Hosted Facts):**
  It is hosted via a neutral or decentralized provider (e.g., IPFS, third-party metadata gateway), specified as private_facts_url in the AgentAddr, enables metadata retrieval without contacting the agent's infrastructure, preserving requester privacy and supporting zero-trust principles and is recommended for agents operating in sensitive or regulated domains.

This dual-hosting model allows agent publishers to choose based on trust requirements, latency preferences, and organizational control.

### D. Resolution Workflow (End-to-End Resolution Workflow with AgentAddr)
The AgentFacts resolution follows a structured, cache-aware workflow:

1. **Index Lookup**
   - **Input:** AgentName (e.g., urn:agent:salesforce:starbucks)
   - **Output:** A signed AgentAddr containing: {agent_id, facts_url, private_facts_url, adaptive_resolver_url (optional), ttl, signature}

2. **Metadata Resolution (Optional)**
   - If trust evaluation or capability validation is needed, the client fetches the AgentFacts from either: facts_url (direct access) or private_facts_url (privacy-preserving access). The AgentFacts is verified using digital signatures and credential chains.

3. **Endpoint Discovery**
   - Clients inspect AgentFacts metadata (if fetched) or AgentAddr (if endpoint is included directly) for:
     - **Static Endpoints**: Stable communication URIs
     - **Rotating Endpoints**: Short-lived, dynamic URIs
     - **Adaptiveresolver URL**: A programmable routing microservice

4. **Endpoint Resolution & Connection**
   - If adaptive_Resolver_url is used, the client sends a request and receives either: A redirect to the optimal endpoint or A signed ephemeral token. Connection to the endpoint is authenticated (e.g., via OAuth2 or JWT) and established.

Clients may choose the resolution strategy based on Latency and load-balancing needs, privacy preferences, credential strength and trust policies, and adaptive behavior triggers. This layered resolution flow allows the system to separate stable identity resolution from dynamic endpoint management, supporting security, agility, and privacy at global scale.

### E. TTL and Caching Strategies
The system uses **time-to-live (TTL)** values to manage caching and resolution frequency across multiple layers: the index (AgentAddr), the agent metadata (AgentFacts), and the dynamic routing layer (e.g., AdaptiveResolver). These TTLs are **signed**, **layer-specific**, and **enforce trust-aware resolution intervals**.

1. **TTL in AgentAddr (Index Layer):** The AgentAddr object returned by the index includes a ttl field. This defines how long the record may be cached by clients, gateways, or edge resolvers before requiring re-validation. It applies to the entire AgentAddr object, including: Metadata pointers (facts_url, private_facts_url) and optional adaptive_resolver_url.

This ensures that the client doesn't frequently burden the index for stable identifiers. TTL values are typically tuned based on the volatility of the agent's state, usage frequency, and trust domain policies.

2. **TTL in AgentFacts (Metadata Layer):** Each AgentFacts includes its own ttl value, often embedded in its cryptographically signed metadata section. This TTL governs: How long the **capabilities**, **telemetry endpoints**, or **evaluations** may be considered fresh and When clients should re-fetch the facts for updated data. This can be used primarily when metadata is used for **trust scoring**, **capability filtering**, or **credential validation**.

3. **TTL for Routing Metadata (Resolution Layer):** Endpoint lists (static or rotating) and adaptive_resolver_url routing logic may have shorter TTLs than AgentFacts metadata. Common TTLs in this layer could be - Static endpoints: 1,6 hours, Rotating endpoints: 5,15 minutes, and Adaptive routing tokens: 30,60 seconds.

TTLs can be adjusted based on agent criticality (e.g., healthcare vs. demo agents), deployment volatility (static hosting vs. serverless endpoints), and trust zone policies.

## VI. ADAPTIVE RESOLUTION AND MULTI-ENDPOINT MANAGEMENT

To meet the demands of globally distributed, privacy-sensitive, and dynamically deployed AI agents, the index architecture supports **adaptive resolution** via programmable, policy-driven components embedded in the agent resolution process. These components decouple **endpoint selection** from static metadata and enable scalable, resilient operation in diverse environments.

### A. Motivation for Adaptive Resolution
Traditional agent endpoint resolution where the agent is tied to a single, static endpoint is inflexible and fragile. AI agents today must serve users from multiple regions with low latency, load balance across backend clusters, obfuscate infrastructure to prevent DDoS attacks and update routing logic in real time without index writes. Adaptive routing solves these challenges by introducing **runtime resolvers** referred to as **AdaptiveResolvers** that inspect request context and forward traffic to the optimal instance.

### B. Multi-Endpoint Resolution Models
Each AgentFacts may contain multiple endpoint references, categorized by their resolution type:
1. **Static Endpoints:** Explicit URIs that represent stable, always-on service interfaces. Ideal for low-frequency, high-trust agents where endpoint volatility is low.

2. **Rotating Endpoints:** A set of URLs with short TTLs, designed for infrastructure that undergoes frequent restarts, redeployments, or geographical rebalancing.

3. **Adaptive Resolver URI:** Points to a microservice or gateway that dynamically routes traffic to optimal downstream instances. Functions as a programmable interface for applying routing logic based on input features (location, time, capability match).
   AgentFacts may define one or more types of endpoints under an endpoints section:

   ```
   "endpoints": {
     "static": ["https://api.example.com/v1"],
     "rotating": ["https://east.example.com", "https://west.example.com"],
     "adaptive_resolver": "https://resolver.example.com/dispatch"
   }
   ```

### C. AdaptiveResolver Component
To avoid requiring AgentFacts resolution for routing in all cases, the index can embed the adaptive_resolver_url directly in the **AgentAddr**, enabling immediate routing without further lookups. Example AgentAddr:

```
{
 "agent_id": ID,
 "adaptive_resolver_url": "https://resolver.example.com/dispatch",
 "ttl": 300,
 "signature": "..."
}
```

**This enables the resolution flow:**

AgentName → Index → AgentAddr → AdaptiveResolver → Ephemeral Endpoint

The **Adaptiveresolver** is an optional but powerful element in the routing stack. It accepts requests, inspects context (e.g., request headers, origin IP, JWT claims), and forwards the call to the most appropriate agent instance. Features include:

- **Load Balancing:** Routes traffic to the least-loaded endpoint instance in real time.
- **Geo-Aware Dispatching:** Selects endpoints closest to the request origin.
- **Capability-Specific Routing:** Directs requests to endpoint versions supporting requested skills (e.g., "real-time translation" vs. "batch mode").
- **Threat Mitigation:** Rotates backend targets or rate-limits sensitive interfaces based on anomaly detection.

The Adaptiveresolver is referenced via the AgentFacts and may itself issue temporary credentials or endpoint tokens to validate downstream access.

**D. Security and TTL Constraints**
All routing logic exposed via the adaptive_resolver_url must **declare TTLs** to define how long routing metadata and endpoint assignments may be cached, **be signed** to prevent manipulation or impersonation and **specify fallback rules**, such as: use static endpoint if resolver is unreachable, retry after a delay with alternate region or elevate to credential-based access.

**E. Implementation Considerations**
For real-world deployments, AdaptiveResolver implementations may take the form:
- **Serverless Functions:** Lightweight routing logic deployed on cloud edge networks.
- **Reverse Proxies (e.g., Envoy, NGINX):** Using programmable rulesets for routing and policy enforcement.
- **Federated Mesh Gateways:** In multi organizations setting, shared resolvers may enforce domain-specific policies.

These implementations must support cryptographic verification, auditability, and availability of SLAs to function reliably within the agent fabric.

**VII. CREDENTIALED TRUST AND PRIVACY MECHANISMS**
The architecture emphasizes **verifiability, privacy, and modular trust** across the entire resolution process. With agents operating in sensitive, autonomous, and federated contexts, all metadata, particularly dynamic endpoint and capability claims must be cryptographically bound and privacy-aware. To support this, the system introduces the separation of:

- **Signed AgentAddr** (returned by the index)
- **Credentialed AgentFacts** (resolved optionally for metadata)
- **Privacy-preserving resolution paths** using third-party infrastructure

Table 6 shows these separated entities and their purpose.

**A. Verifiable Credentials and Signed Metadata**
The integrity of agent metadata and its update mechanism depends on **W3C Verifiable Credentials (VCs)** to assert capabilities, skills, audits, or evaluations and **cryptographic signatures** to bind these claims to the agent identity. This ensures agents cannot spoof skills, impersonate trusted services, or hijack reputational pathways.

| Artifact | Purpose | Signed By |
| --- | --- | --- |
| AgentAddr | Identity, routing pointers, TTL | Index resolver |
| AgentFacts | Capabilities, credentials, endpoints | Credential issuers |
| Endpoint Tokens | Temporary routing/session bindings | AdaptiveResolver (optional) |

**Table 6:** Artifact within an entry on the NANDA index architecture, and their purpose

**B. Credential Governance and Trust Domains**
The trustworthiness of an AgentFacts hinges on the validity and recognition of its credential issuers. Our system supports a **federated trust governance model**, where different credential authorities may govern their own "trust zones" (e.g., within an enterprise, vertical, or region). These zones can define credential schemas and issuance policies, establish revocation mechanisms, and inter-operate using cross-signing, federation agreements, or index (registry)-to-index (registry) bridges. Clients interacting with agents may configure their own trust preferences - accept only credentials from whitelisted authorities, require threshold verification (e.g., at least two independent audits), and verify issuer reputation using embedded TRS (Trust Reputation Scores).

This modular design allows the ecosystem to evolve without hardcoded trust anchors.

### C. Privacy-Preserving Resolution via PrivateFactsURL

To protect the **privacy of the accessor**, the system supports resolving agent metadata via a third-party or decentralized location (e.g., IPFS, secure proxy), specified as private_facts_url in the AgentAddr. This model provides

1) **Requester anonymity**: Agent domain is never contacted
2) **Access decoupling**: Metadata is publicly readable without invoking agent infrastructure
3) **Audit independence**: Metadata remains available even if the agent hosting goes offline

Clients may be configured to **prefer this route** when operating in regulated or high-sensitivity environments.

AgentName → Index → AgentAddr → PrivateFactsURL → AgentFacts (metadata only)

### D. Revocation and Credential Freshness

Credential freshness is essential for safety in real-time environments. To support dynamic trust, the system includes:
- **Credential Expiry Fields:** Embedded in all signed objects.
- **Revocation Lists:** Served by credential issuers and queried by clients during interaction.
- **Hash-Linked Chains:** Credential validity can be chained across issuers (e.g., capability certified by org A, endorsed by org B).

It is advised to recompute TRS on each new session, set cache expiration policies shorter than the weakest credential's TTL, and perform multi-source validation for high-risk agents (e.g., financial or healthcare applications).

### E. Resolution Layer Enforcement

The architecture enforces that AgentAddr records can only be resolved and cached within their TTL window, AgentFacts metadata is not assumed to be static; clients must verify signatures and freshness, and privacy-preserving resolution (via PrivateFactsURL) is functionally equivalent to FactsURL, but adds a layer of **access decoupling**

### VIII. DEPLOYMENT CONSIDERATIONS AND USE CASES

This section outlines practical implementation strategies for deploying the proposed index architecture and highlights real-world use cases that demonstrate its applicability across domains. The architecture's lean index, dynamic resolution, and credentialed metadata design make it well-suited to diverse operational environments ranging from tightly governed enterprise systems to open agent marketplaces.

### A. Deployment Models

The index and AgentFacts system can be deployed under multiple architectural models, depending on organizational goals, control boundaries, and compliance requirements.

1. **Enterprise-Controlled Registries**
   - Enterprises deploy and manage internal registries governing their agent namespaces.
   - AgentFacts are hosted within their infrastructure (PrimaryFactsURL).
   - Credential authorities reside within the enterprise's compliance and audit teams.

2. **Federated Industry Registries**
   - Vertically integrated industries (e.g., logistics, healthcare) operate consortium registries.
   - AgentFacts reference third-party capabilities and certifications from accredited bodies.
   - Governance rules are encoded in index policy and enforced via signature validation.

3. **Decentralized Public Registries**
   - Publicly accessible registries allow open registration of agents.
   - Credential validation and revocation are enforced via smart contracts or decentralized protocols.
   - AgentFacts are hosted in IPFS or similar decentralized file systems, emphasizing privacy and censorship resistance.

4. **Hybrid Topologies**
   - Organizations may register with both private and public registries for broader discoverability.

- Trust rules and resolution pathways allow clients to prioritize index sources based on use-case-specific logic.

These models enable the system to adapt to both regulated, closed-loop deployments and open, innovation-driven ecosystems.

**B. Performance and Scaling Considerations**

While the index is intentionally lean, the supporting infrastructure (AgentFacts hosts, adaptive resolvers, credential resolvers) must be designed for internet-scale agent traffic. Key strategies that could be helpful:
- **Edge Caching:** TTL-enforced caching of AgentFacts and routing metadata at CDN edges to reduce resolution latency.
- **Sharded Credential Verification:** Parallel trust scoring engines operating at domain-specific points of presence (PoPs).
- **Load-Adaptive Resolvers:** Stateless microservices that scale horizontally and resolve endpoints dynamically based on request volume and client location.
- **Audit and Monitoring Hooks:** Integration with OpenTelemetry and observability stacks to support SLA monitoring and behavior-based trust score computation.

## IX. CONCLUSION

The rise of autonomous AI agents presents unprecedented demands on digital infrastructure, scalable identity, verifiable trust, dynamic coordination, and privacy-preserving discovery. Traditional internet protocols, built for static, human-operated endpoints, are insufficient for the fluid, high-volume interactions required by agent-scale autonomy. This paper introduced a lean, modular index architecture as a foundational layer for the decentralized Internet of AI Agents. Grounded in the NANDA Protocol, the design advances the state of the art across multiple dimensions:

- **Minimal Core Index**: Decouples static identifiers from dynamic metadata, reducing update frequency, exposure, and attack surfaces.
- **Credentialed AgentFacts**: Enable rapid, verifiable updates to agent capabilities, routing logic, and telemetry.
- **Timed Resolution & Adaptive Routing**: Support latency optimization, DDoS mitigation, and geo-aware deployments.
- **Privacy-Preserving Discovery**: Protects accessor identities and aligns with zero-trust security models.
- **Federated Trust & Governance**: Enables scalable oversight across jurisdictions, organizations, and policy domains.

Together, these components establish the groundwork for an open, secure, and extensible agent ecosystem capable of supporting trillions of agents operating cooperatively across digital and physical environments.

## X. FUTURE WORK

Several avenues for future research and development build upon this foundation:
1. **Trust Layer Extensions**
   Future work will explore decentralized trust score computation, integration with behavioral monitoring systems, and real-time reputation feeds, inspired by directions highlighted by [16].
2. **Governance Frameworks and Revocation Models**
   We will formalize revocation protocols, dispute resolution mechanisms, and multi-jurisdictional governance workflows to enforce trust at scale. This includes modeling hierarchical and mesh-based governance for industry-specific applications.
3. **Open Standards and Interoperability**
   Standardizing the AgentFacts schema with IETF/W3C bodies and enabling interoperability with Open Agent Frameworks, and existing Web3 identity platforms will ensure ecosystem adoption.
4. **Privacy Enhancements with Zero-Knowledge Proofs (ZKPs)**
   To further privacy guarantees, ZKP-based credential assertions [17] can be introduced, allowing agents to prove capabilities or compliance without revealing full certificate chains.

## XI. OPEN QUESTIONS AND DISCUSSION

While this paper proposes a comprehensive index and metadata architecture for AI agents, several key areas remain open for further exploration, community alignment, and empirical validation. These questions span technical design, trust models, privacy trade-offs, economic incentives, and future governance structures. Their resolution will shape the long-term viability and adaptability of decentralized agent infrastructure at a global scale.

**A. Hosting and Accessibility of Agent Metadata**

A central question concerns the hosting model for AgentFacts. The current design permits multiple options: (i) self-hosting by the agent (e.g., at /.well-known), (ii) hosting through secure third-party services, or (iii) provisioning by a neutral infrastructure provider such as NANDA. Each approach offers distinct trade-offs in privacy, trust, and update latency.

- Self-hosting ensures provenance but reveals access patterns.
- Third-party hosting enhances privacy but introduces a dependency on the host's security posture.
- Public infrastructure democratizes access but may incur governance and scalability concerns.

A flexible model is desirable, but the trade-offs between decentralization, performance, and control remain unresolved.

**B. Index Visibility and Discoverability**

Unlike DNS, the agent index may be intentionally opaque to prevent abuse or sensitive agent enumeration. At the same time, open discovery is crucial for enabling federation, crawling, and reputation-based search.

**Open issue**: Should the index support machine-readable scraping or indexing, and if so, under what access controls or throttling policies?

This question has profound implications for building agent reputation graphs, public directories, or sector-specific trust indexes.

**C. Terminology Alignment with Prior Art**

The use of the term *AgentFacts* may overlap or cause confusion with prior frameworks such as A2A. While our schema significantly extends prior work, we may consider renaming it to *Agent Metadata Layer* or *AgentFacts v2*, and clearly describing its relationship to A2A.

Additionally, terms like PrimaryFactsURL and PrivateFactsURL may be refined to more functional descriptors such as *Primary Metadata Endpoint* and *Privacy-Preserving Metadata Endpoint*.

**Open issue**: What terminology best communicates extensibility while acknowledging lineage and compatibility with existing systems?

**D. Resolution Flow and Return Semantics**

Current resolution logic implies that accessing an endpoint requires first resolving and validating the AgentFacts. However, it is unclear whether the client must always parse the facts or if a signed *AgentAddr* or *AgentLoc* object containing endpoint and metadata digests would suffice.

**Open issue**: Should the resolver return a lightweight signed object, distinct from the full facts, optimized for runtime communication?

This separation could reduce latency and support pluggable authentication protocols at the edge.

**E. TTL Strategy and Update Constraints**

While the paper introduces TTL-based caching, important design questions remain:

- Who determines TTL values agents, registries, credential authorities, or external policy engines?
- Should TTLs vary by agent type (e.g., real-time agents vs. scheduled agents)?
- Can TTLs be **ephemeral** or **task-conditioned**, based on state machines or mission duration?

**Open issue**: How do we strike a balance between agility and stability in TTL assignments, particularly for high-frequency or regulated agents?

**F. Credential Governance and Delegation**

Another unresolved dimension is the delegation model for updates and credential issuance:

- Can agents update their own metadata autonomously?
- Are delegated authorities (e.g., platform vendors or consortia) responsible for publishing credentials?
- Can agents authorize credentialing-on-the-fly via smart contracts or concierge agents?

**Open issue**: What are the minimal viable trust constraints to allow automated updates while preventing spoofing, abuse, or trust inflation?

**G. Adaptive Routing Reliability**

The adaptive routing layer introduces flexibility and performance, but it also poses new failure modes:

- What fallback paths should be employed if the AdaptiveResolver is offline?
- Can routing logic be cached or prevalidated in critical systems?
- How is routing evaluated for fairness and performance across federated domains?

**Open issue**: What guarantees must the resolution and routing layer provide to support mission-critical or regulated agent deployments?

**H. Economic Incentives and Monetization Models**

The sustainability of agent registries and metadata services may depend on viable economic incentives. Yet the architecture must avoid rent-seeking behavior or exclusionary gatekeeping.

**Open issue**: Should AgentFacts hosting or TTL adjustments be linked to micropayments, staking mechanisms, or usage tiers? How can such models preserve openness and composability?

**APPENDIX**

**Full AgentFacts Schema**
Fields marked with blue are new to AgentFacts, while fields in green are present in Google AgentCard. The AgentFacts schema here points to a Salesforce-native translation agent that works according to the NANDA schema.

```
{

  // ========== IDENTITY & BASIC INFORMATION ==========

  // 🔵 Unique machine-readable
  "id": "nanda:550e8400-e29b-41d4-a716-4466554400",

  // 🔵 Agent URN identifier
  "agent_name": "urn:agent:salesforce:TranslationAssistant", //Can be traditional, or can be DID: in this case, a Salesforce native agent

  // 🟢 Human readable name (maps to AgentCard.name)
  "label": "TranslationAssistant",

  // 🟢 Agent description (maps to AgentCard.description)
  "description": "Autonomous agent for low-latency multilingual translation",

  // 🟢 Version information (maps to AgentCard.version)
  "version": "1.2.1",

  // 🟢 Documentation URL
  "documentationUrl": "https://docs.salesforce.com/agents/translation",

  // 🔵 Jurisdiction: which country/entity covers the compliance for this agent
  "jurisdiction": "USA",

  // ========== PROVIDER INFORMATION ==========

  // 🟢 Provider details (maps to AgentCard.provider)
  "provider": {
    "name": "Salesforce",
    "url": "https://salesforce.com",
    // 🔵 Optional Decentralized identifier for provider verification
    "did": "did:web:salesforce.com"
  },

  // ========== NETWORK ENDPOINTS ==========

  "endpoints": {
    // 🟢 Static endpoints (maps to AgentCard.url)
    "static": [
      "https://api.salesforce.com/v1/translation" //Can be a traditional endpoint or a DID-based endpoint
    ],
    // 🔵 Dynamic routing capabilities
    "adaptive_resolver": {
      "url": "https://resolver.salesforce.com/dispatch/translation",
      "policies": [
```

```json
        "geo",            // Geographic routing
        "load",           // Load balancing
        "threat-shield"   // Security policies
      ]
    }
  },

  // ========== TECHNICAL CAPABILITIES ==========

  "capabilities": {
    // 🟢 Input/output modalities (maps to AgentCard.defaultInputModes/defaultOutputModes)
    "modalities": [
      "text",
      "audio"
    ],

    // 🔵 Real-time streaming support
    "streaming": true,

    // 🔵 Batch processing support
    "batch": false,

    // 🟢 Authentication methods (maps to AgentCard.securitySchemes & security)
    "authentication": {
      "methods": [
        "oauth2",
        "jwt"
      ],
      "requiredScopes": [
        "translate:real-time",
        "language:detect"
      ]
    }
  },

  // ========== FUNCTIONAL SKILLS ==========

  // 🟢 Skills array (maps to AgentCard.skills)
  "skills": [
    {
      // 🟢 Skill identifier
      "id": "translation",

      // 🟢 Skill description
      "description": "Real-time translation between >25 languages",

      // 🟢 Input modes for this skill
      "inputModes": [
        "text",
        "audio/ogg"
      ],

      // 🟢 Output modes for this skill
```

```json
      "outputModes": [
        "text",
        "audio/wav"
      ],

      // 🔵 Language support specification
      "supportedLanguages": [
        "en", "es", "fr", "de", "ja", "zh"
      ],

      // 🔵 Performance constraint
      "latencyBudgetMs": 300
    },
    {
      "id": "summarisation",
      "description": "Topic-guided abstractive summarisation",
      "inputModes": ["text"],
      "outputModes": ["text"],
      // 🔵 Token limits
      "maxTokens": 2048
    },

  ],

  // ========== QUALITY METRICS ==========

  // 🔵 Certified performance and reliability metrics
  "evaluations": {
    "performanceScore": 4.8,
    "availability90d":"99.93%",
    "lastAudited": "2025-04-01T12:00:00Z",
    "auditTrail": "ipfs://QmX7Y8Z9...",   // Immutable audit evidence
    "auditorID": "Capabilities Auditor v2.1" }
  },

  // ========== OBSERVABILITY & MONITORING ==========

  // 🔵 Telemetry and monitoring configuration
  "telemetry": {
    "enabled": true,
    "retention": "7d",
    "sampling": 0.1,
    "metrics": {
      "latency_p95_ms": 280,
      "throughput_rps": 125,
      "error_rate": 0.003,
      "availability": "99.93%"
    }
  },

  // ========== TRUST & VERIFICATION ==========

  // 🔵 Certification and trust framework
```

```
  "certification": {
    "level": "verified",
    "issuer": "Salesforce",
    "issuanceDate": "2025-03-15T09:30:00Z",
    "expirationDate": "2026-03-15T09:30:00Z"
  },
}
```